\documentclass[twocolumn]{aastex63}

\shorttitle{Radio, EUV, and X-Ray Observations During a Filament Rise}
\shortauthors{Karlick\'y, Ka\v{s}parov\'a, and Sych}
\begin{document}
\received{2019 August 21}
\revised{2019 November 12}
\accepted{2019 November 13}

\title{Radio, EUV, and X-Ray Observations During a Filament Rise in the 2011 June 7 Solar Flare}

\author[0000-0002-3963-8701]{Marian Karlick\'{y}}
\affiliation{Astronomical Institute of the Czech Academy of Sciences, Fri\v{c}ova 298, Ond\v{r}ejov, 251 65, Czech Republic}

\author[0000-0001-9559-4136]{Jana Ka\v{s}parov\'{a}}
\affiliation{Astronomical Institute of the Czech Academy of Sciences, Fri\v{c}ova 298, Ond\v{r}ejov, 251 65, Czech Republic}

\author[0000-0003-4693-0660]{Robert Sych}
\affiliation{Institute of Solar-Terrestrial Physics SB RAS, Irkutsk, Russia}

\correspondingauthor{M. Karlick\'{y}}
\email{karlicky@asu.cas.cz}

\begin{abstract}
The most energetic flares start with a filament rise followed by magnetic
reconnection below this filament. The start of the reconnection corresponds to
the beginning of the flare impulsive phase. In this paper we study processes
before this phase. During the filament rise we recognize an
unusual radio continuum with a starting boundary drifting toward lower
frequencies. The estimated velocity of the agent generating this continuum
boundary is about 400 km s$^{-1}$, similar to that  of the
rising filament. In association with this filament rise, transient X-ray
sources and extreme ultaviolet (EUV) brightenings are found near the filament footpoint and
outside the locations where later two parallel flare ribbons appear.  Moreover,
oscillations with a $\sim$30 s period are found simultaneously in radio, 
EUV, and X-ray observations. Around the end of these oscillations the flare
impulsive phase starts as seen in observations of the drifting pulsation
structure and X-ray source located at the upper part of the rising filament. We
interpret the unusual radio continuum and transient X-ray sources, which are
located outside the two parallel flare ribbons, as those generated during an
interaction of the rising filament with the above-lying magnetic loops. The
EUV brightening at the filament footpoint could be a signature of the
magnetic reconnection inside the magnetic rope carrying the filament.
Possible scenarios of the $\sim$30 s period oscillations in radio,
X-ray, and EUV are discussed.
\end{abstract}


\section{Introduction}

As expressed in a three-dimensional model of two-ribbon flares, these flares start
with a filament eruption \citep{2012A&A...543A.110A,2013A&A...549A..66A}.
While the evolution of such flares during and after their impulsive phase has been
studied in many papers \citep[for review see, e.g.,][]{2011SSRv..159...19F}, the
very early phase just after the start of the filament rise and before the impulsive
phase is still not well understood. This is due to the fact that this
phase is usually relatively short. However, in the 2011 June 7 flare associated
with the ejection of an unusually huge filament, this pre-impulsive phase was
relatively long, i.e. nearly 10 minutes. This flare had already been studied in
several papers. For example, \cite{2012ApJ...745L...5C} studied the formation
of extreme ultraviolet (EUV) wave from the expansion of a coronal mass ejection (CME). They found that,
following the solar eruption onset, the CME exhibits a strong lateral expansion
and the expansion speed of the CME bubble increases from
100 km s$^{-1}$ to 450 km s$^{-1}$ in only six minutes. 
\cite{2013ApJ...777...30I} studied motions of the X-ray and UV sources
at times starting with the flare impulsive phase at 06:25 UT. They described
substantial parallel and perpendicular motion of the hard X-ray footpoints. The
motion of the footpoints parallel to the flare ribbons is unusual; it reverses
direction on at least two occasions. \cite{2014ApJ...788...85V} investigated
the coronal magnetic reconnection driven by the CME. They
presented observations together with a data-constrained numerical simulation,
demonstrating the formation/intensification of current sheets along a
hyperbolic flux tube at the interface between the CME and the neighboring
active region. \cite{2016ApJ...827..151Y} studied a flux cancellation during
the filament eruption. There are also the papers describing filament blobs in
this filament eruption, especially those falling back to the
Sun~\citep{2012A&A...540L..10I,2016A&A...592A..17I,2013ApJ...776L..12G,2014ApJ...782...87C}.

The main purpose of this paper is to investigate in detail the relatively long
pre-impulsive phase of the 2011 June 7 flare (SOL2011-06-07T06:16). We present
an unusual radio continuum observed at this phase together with
transient X-ray sources and the oscillations, recognized simultaneously in
hard X-ray, EUV, and radio emissions. For analysis of these oscillations
we use not only the well known wavelet power spectra
\citep[see, e.g.,][]{2015ApJ...798....1I,2016ApJ...827L..30H}, but also the wavelet
amplitude spectra. While the wavelet power spectrum expresses the
square of the absolute value of the wavelet coefficients, the wavelet
amplitude spectrum corresponds to the real part of the wavelet
coefficients
\citep{1996PhyU...39.1085A,2002ChJAA...2..183S,2008SoPh..248..395S}. Although
the wavelet amplitude spectra are not frequently used by the solar physics
community, we consider them to be very useful because they show not only periods
but also phases of the oscillations. We also use these spectra for an
analysis of the relationship between different light curves, applying the
wavelet coherence and wavelet cross-correlation digital techniques. Microwave
observations from Nobeyama radioheliograph and radio polarimeters are also analyzed.
Finally, all observations are discussed and a scenario for the
pre-impulsive flare phase is suggested.

\begin{figure}[t]
\begin{center}
\includegraphics*[width=0.48\textwidth]{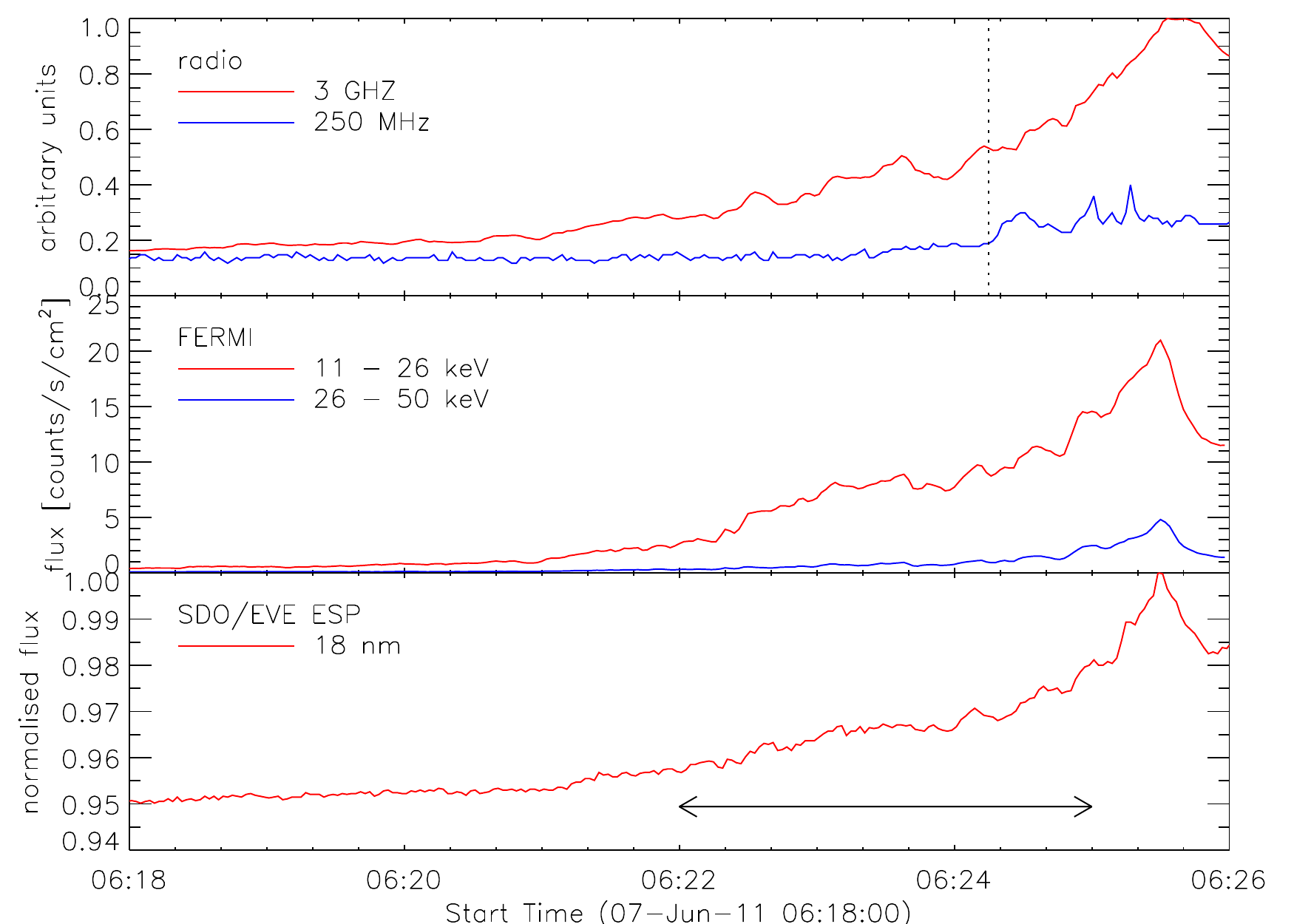}
\end{center}
\caption{Light curves of \textit{Fermi}/GBM detector No. 5, EVE/ESP 18 nm, and radio
  emission at 3 GHz and 250 MHz.
  For display purposes, these data were rebinned to a 2s time resolution.
  The vertical line in radio denotes
  the start of the drifting pulsation structure (start of the flare impulsive phase).
  The horizontal line with arrows indicates the interval of the $\sim$ 30 s oscillations, approximately.
}
\label{light_curves}
\end{figure}
\section{Observations and  Analysis}

In the active region NOAA 11226, on 2011 June 7, an M2.5 class flare was
observed at the location of S22$^{\circ}$W53$^{\circ}$. It started at 06:16 UT
with the eruption of a massive filament.

The light curves of the pre-impulsive phase of this flare in X-rays \citep[\textit{Fermi}
Gamma-Ray Burst Monitor (\textit{Fermi}/GBM);][]{2009ApJ...702..791M}, EUV
\citep[EUV Experiment (EVE/ESP);][]{2012SoPh..275..115W} and radio (3 GHz radiotelescope;
\citealt{1993SoPh..147..203J}; BLEN7M-Callisto; \citealt{2005SoPh..226..143B} at
250 MHz) are shown in Fig.~\ref{light_curves}. As can be seen, the
emission in all bands increases up to the first peak at 06:25:30 UT and the
light curves show quasi-periodic pulsations. The impulsive phase in radio
emission started at about 06:24:15 UT as shown by the vertical dashed line.

\begin{figure}[!t]
\begin{center}
\includegraphics*[width=0.48\textwidth]{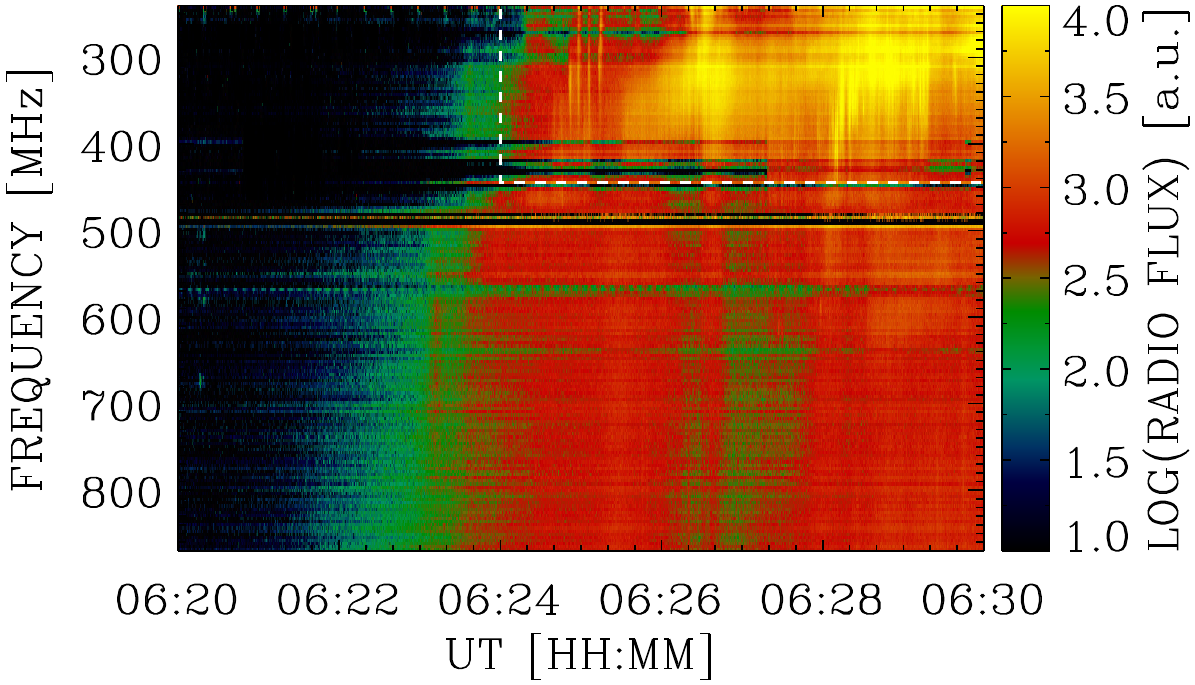}
\includegraphics*[width=0.48\textwidth]{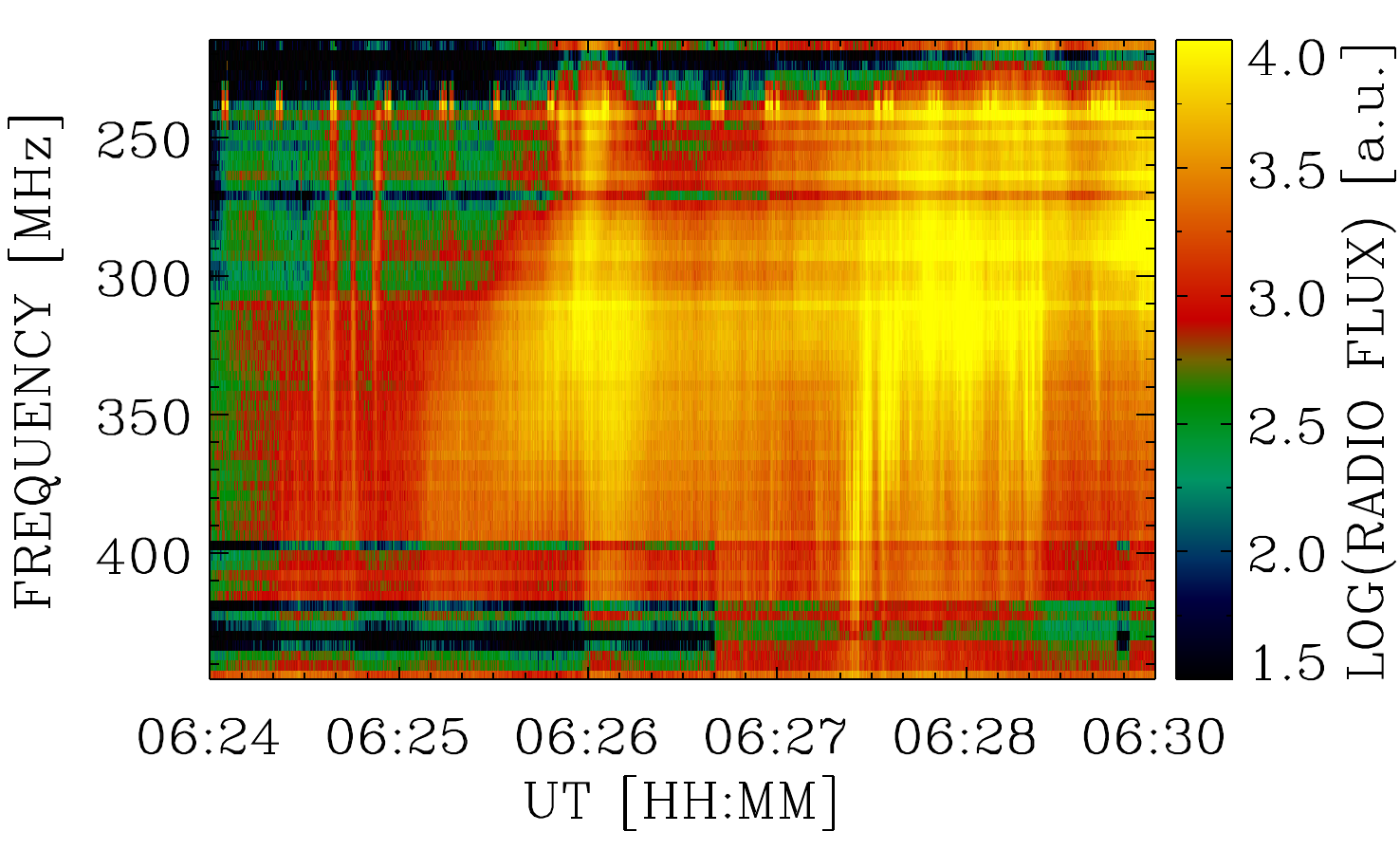}
\end{center}
    \caption{Top: BLEN7M-Callisto radio spectrum observed on 2011 June 7,
    showing an unusual continuum which starts at 870 MHz at about 06:21:40 UT
      and whose starting boundary drifts from high to low frequencies.
    This emission is followed by the drifting pulsation structure (DPS) observed in the 220--450 MHz
    range at about 06:24:15--06:30 UT. Bottom: zoom of the spectrum with DPS, corresponding to the box delineated by
    the dashed line in the top panel.}
\label{blen}
\end{figure}

\subsection{Radio observations}

Figure~\ref{blen} displays the BLEN7N-Callisto radio spectrum of the beginning
of the 2011 June 7 flare \citep{2005SoPh..226..143B}. (We note that a similar
spectrum was also observed by OSRA-Callisto at Ond\v{r}ejov observatory.)
As shown, the radio emission starts with the continuum emission
whose starting boundary drifts from high to low frequencies. The continuum
starts at 870 MHz at about 06:21:40 UT, but at 300 MHz it starts at about 06:24:20 UT.
Thus, the mean drift of the starting continuum boundary is about -3.6 MHz
s$^{-1}$. This continuum is unusual. To our knowledge, a similar
emission was only observed at the beginning of the 2001 September 24 flare (see
Fig. 14 in the paper by ~\citealt{2003A&A...399.1159F}), in association with a loop
ejection. But the radio emission from that flare was observed
at lower frequencies and with a lower frequency drift, $\sim$ -0.1 MHz~s$^{-1}$, than in the present case.

\begin{figure}[t!]
\begin{center}
\includegraphics*[width=0.49\textwidth]{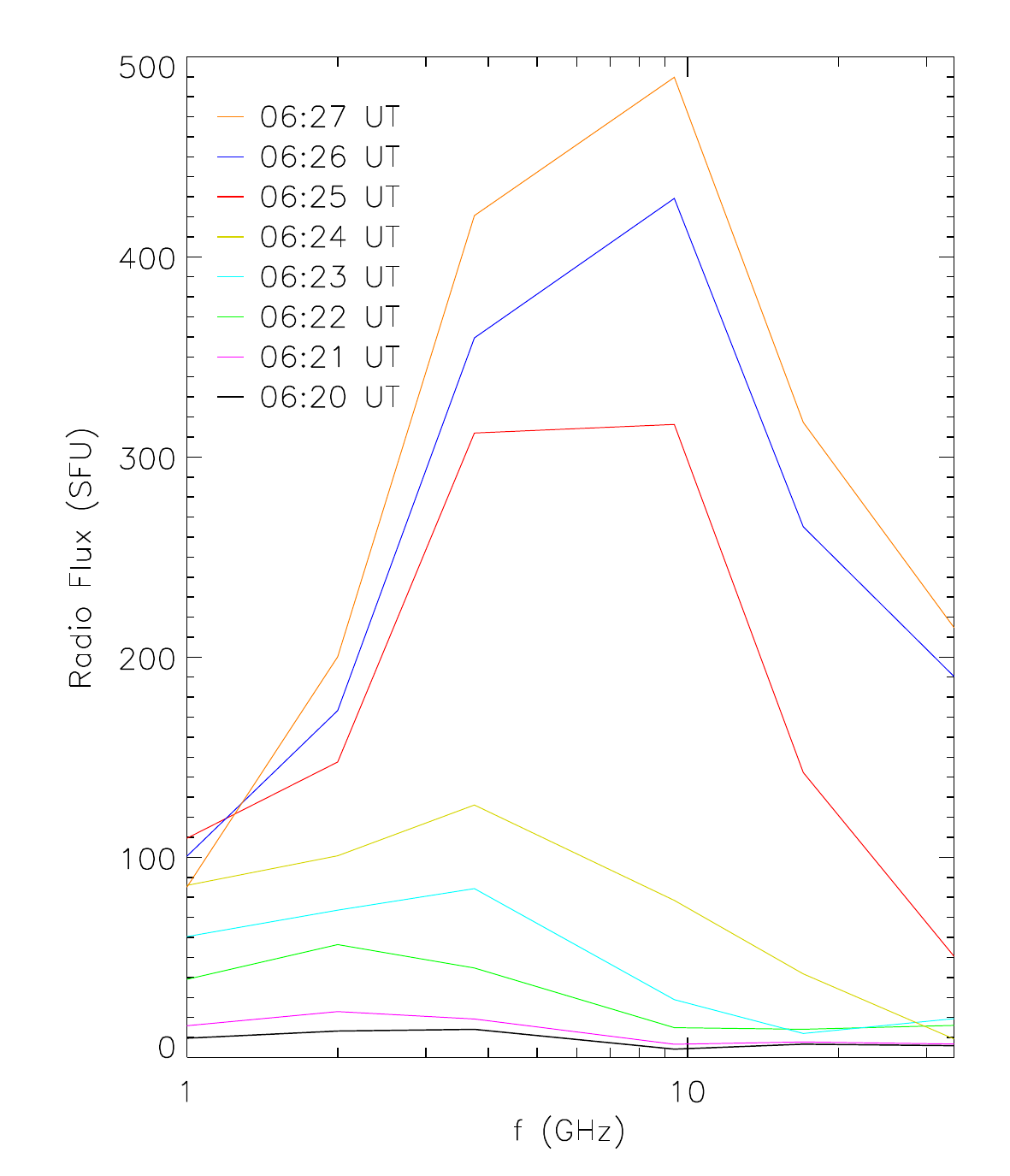}
\end{center}
    \caption{Spectra of the flare microwave emission observed by the Nobeyama radio polarimeters from
    06:20 UT to 06:27 UT.}
\label{micro}
\end{figure}

Fig.~\ref{blen} and its zoom show that this unusual radio
continuum is followed by the drifting pulsation structure (DPS), observed in
the 220--450 MHz range and at about 06:24:15--06:30 UT. It consists of a
series of pulses that as a whole drifts to lower frequencies. DPS is proposed
to be a signature of the plasma emission from the plasmoid located in the
rising magnetic rope~\citep{2000A&A...360..715K,2002A&A...395..677K}.
Individual pulses in the DPS are generated by superthermal electrons
quasi-periodically accelerated in the magnetic reconnection below the rising
filament and trapped in the plasmoid. Because the plasmoid is moving upward in
the solar atmosphere, its plasma density decreases and thus the DPS drifts to lower
frequencies. We note that this DPS is observed at very low frequencies;
usually, they are observed in the GHz frequency
range~\citep{2015ApJ...799..126N}.

This flare was also observed by the Nobeyama radioheliograph at 17 and 34 GHz
\citep{1994IEEEP..82..705N} and by the Nobeyama radio polarimeters at six single
frequencies: 1, 2, 3.75, 9.4, 17, and 35 GHz \citep{1985PASJ...37..163N}. The
evolution of the 17 and 34 GHz sources together with the spectral index is
shown in a movie available at the Nobeyama radioheliograph page for this
event\footnote{\url{http://solar.nro.nao.ac.jp/norh/html/event/20110607_0627/norh20110607_0627.html}}.
As can be seen there, the 17 and 34 GHz sources at the position [700,
  -370] arcsec are extending in time.

Figure~\ref{micro} shows microwave spectra of this flare obtained from the
Nobeyama radio polarimeters data at time 06:20--06:27 UT. At this frequency
range the emission is generated by a gyro-synchrotron mechanism. These
spectra show a rapid change between 06:24 and 06:25~UT, which
corresponds to the start of the flare impulsive phase as indicated also by the
start of the DPS.

\subsection{X-Ray and EUV Observations}

We also studied data from \textit{RHESSI} \citep{2002SoPh..210....3L} and \textit{SDO}/AIA
\citep{2012SoPh..275...17L}. \textit{RHESSI} X-ray sources show complex structure and
time evolution on a time scale of
  several seconds. Therefore, we used
  {\it Clean} and {\it VIS\_CS} \citep{2017ApJ...849...10F} algorithms for X-ray source reconstruction
  to check hard X-ray (HXR) source positions. The reliability of the image reconstruction was
  assessed by comparing observed and expected (from the reconstructed image)
  modulation and visibility profiles, available through the \textit{RHESSI} software. In most cases,
  both algorithms produced similar results in terms of reconstructed source positions
  and agreement of the observed and image profiles.
  However, if the image structure differed significantly between the two algorithms,
  e.g. if HXR sources were not located
  within 50\% level of the maximum flux, the {\it Pixon} \citep{1996ApJ...466..585M} algorithm was used as a further check.
  Other image reconstruction algorithms available in the \textit{RHESSI} software
  did not produce comparably good agreement of the observed and image profiles.
  To account for the HXR sources' time evolution, the time interval for the image reconstruction was
  adjusted so that the total number of counts of the reconstructed image was at least $10^3$.
  HXR images were reconstructed in three energy channels: 6--12, 12--25, and 25--50 keV, with
  the time interval ranging from 28 to 12~s using front grids 4--9.
  HXR sources reconstructed by {\it Clean} tend to be larger than those obtained by other
  algorithms; we did not remove the \textit{RHESSI} instrument point spread function and used the {\it Clean} default version; see
  also \citet{2009ApJ...698.2131D}.
  On the other hand, {\it VIS\_CS} caused fragmentation of the image in few cases or resulted
  in artificial sources, i.e. strong sources which were not reconstructed by  {\it Clean} or {\it Pixon};
  see e.g. Figure~\ref{x0622}. Generally, {\it Clean} image modulation profiles  showed better agreement with
  the observed ones for the coarsest (8 and 9) \textit{RHESSI} grids, whereas {\it VIS\_CS} resulted
  in better agreement for 4--6 grids both in modulation and visibility profiles.

\begin{figure}[t]
\begin{center}
\includegraphics*[width=0.49\textwidth]{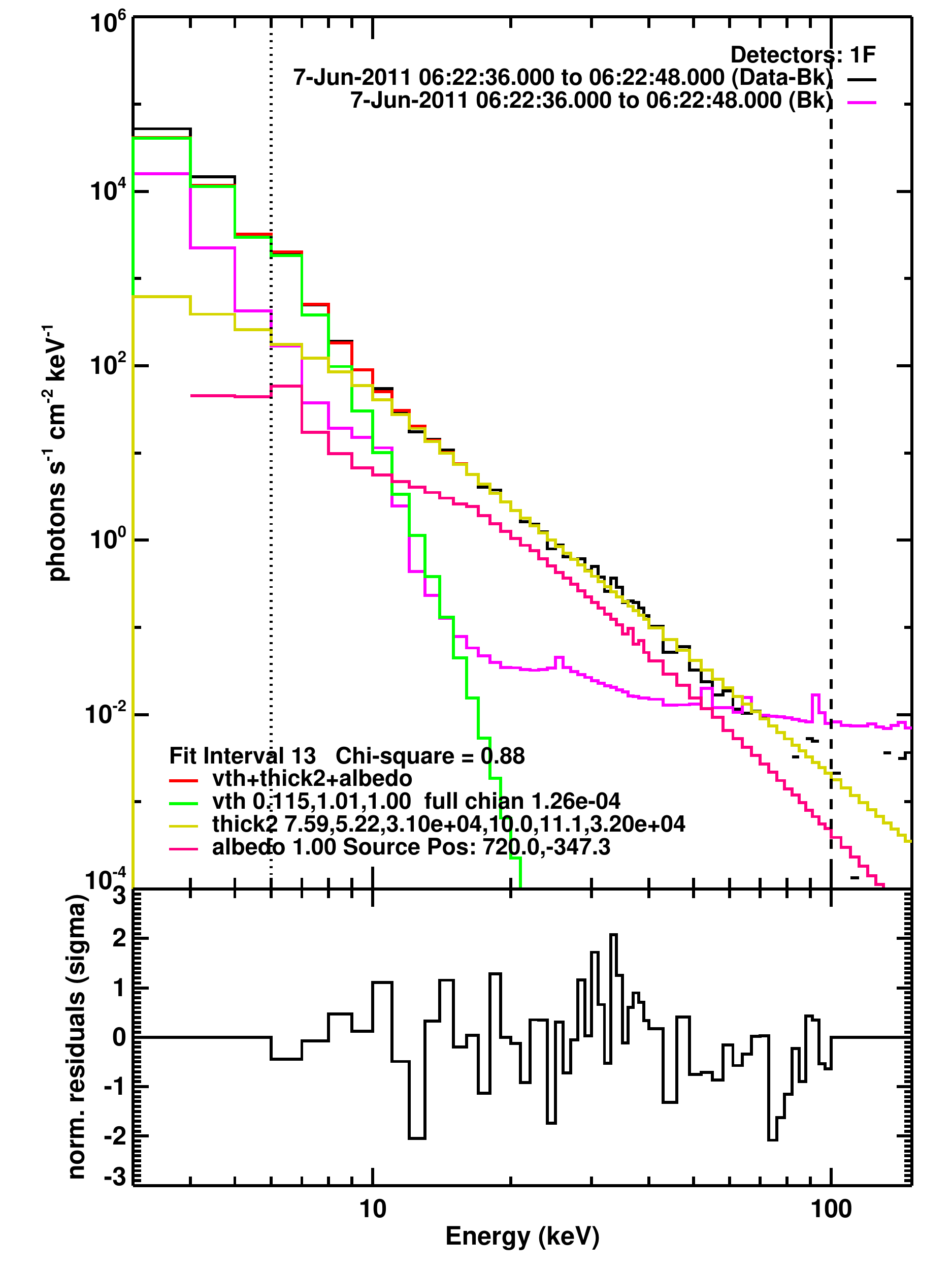}
\end{center}
    \caption{
      \textit{RHESSI} photon spectrum in the 06:22:36--06:22:48 UT time interval fitted with isothermal, albedo, and
      thick-target components. In the thick-target component the break energy and power-law index above it
      were fixed to use a single power-law function for this component.
      The bottom plot shows normalized residuals. Vertical lines denote the
    energy range used in the fitting.}
\label{xspec0622}
\end{figure}

    Spatially integrated spectra were analyzed using front detectors 1, 3, 4, 5, 6, 8, and 9
  in the 12s time intervals from 06:20:00 UT. Background subtracted spectra were fitted with
  isothermal, albedo, and thick-target components. The non-thermal component dominated X-ray
  spectrum above $\sim$10~keV, the low-energy cutoff gradually increased from $\sim$10~keV at
  06:20 UT to $\sim$15~ keV at 06:26 UT.
  An example of the fitted \textit{RHESSI} spectrum from detector 1 corresponding to the time interval
  of the \textit{RHESSI} image  in Figure~\ref{x0622} is shown in Figure~\ref{xspec0622}.
  Therefore, we consider all HXR sources above $\sim$~12~keV as
  non-thermal ones produced by accelerated particles.

  \textit{SDO}/AIA images were obtained from two sources. First, \textit{SDO}/AIA cutouts
    were available via {\tt show\_synop.pro} interface in SolarSoftWare. These data were not modified.
    Second, to achieve a higher time resolution in some AIA filters,
    level 1 data from the JSOC\footnote{\url{http://jsoc.stanford.edu/}} interface were used as well. These were corrected for stray light
    using the method of \citet{2013ApJ...765..144P}.

\begin{figure}[t]
\begin{center}
\includegraphics*[width=0.48\textwidth]{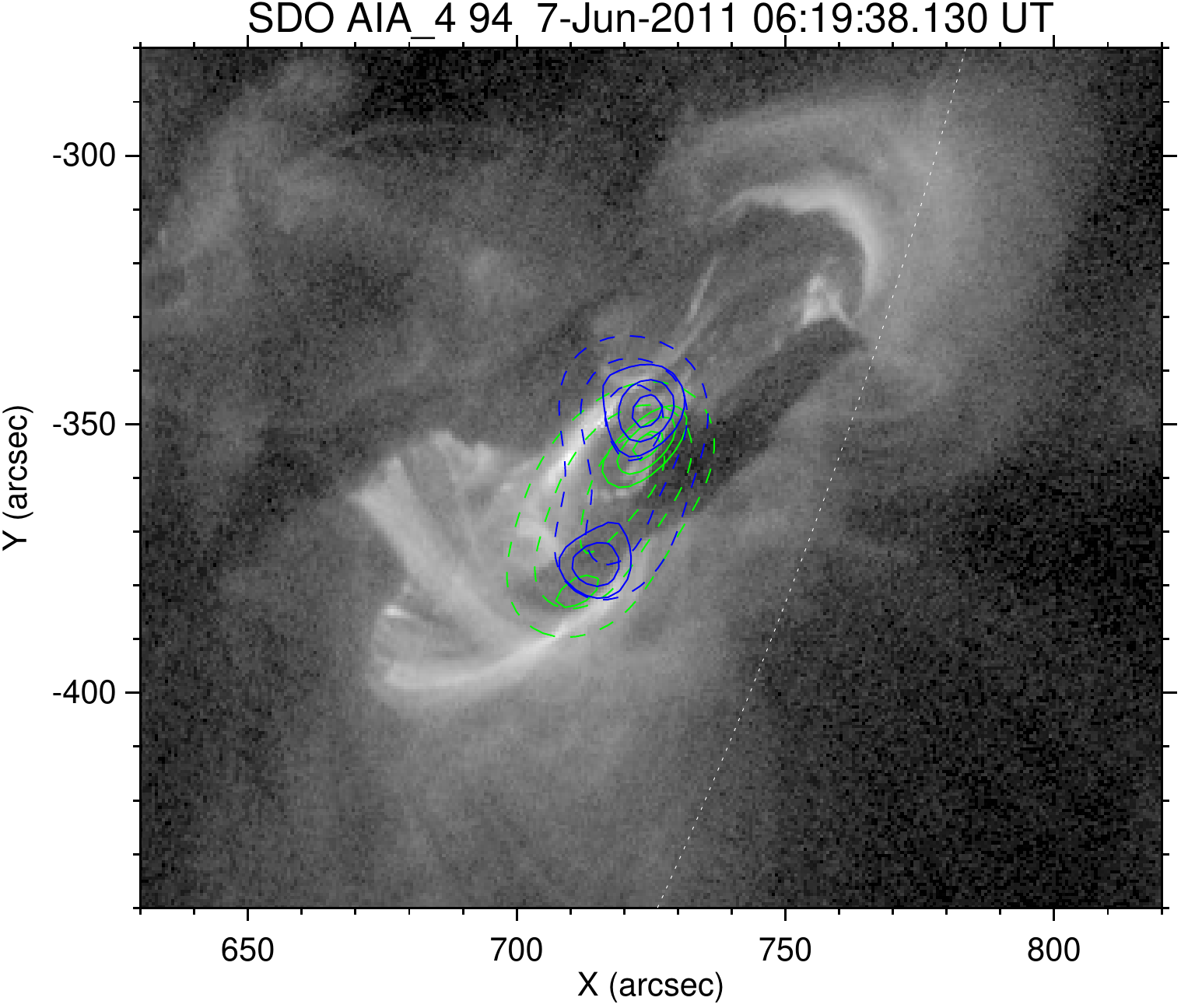}
\end{center}
    \caption{
    \textit{RHESSI} X-ray sources in the 6--12~keV (green) and 12--25~keV (blue) energy ranges in the
      06:19:36--06:19:48 UT time interval  superposed on the \textit{SDO}/AIA 94~\AA~image at 06:19:38 UT.
      Contours correspond to 50, 70, and 90\% levels of the maximum in the
      images. Full lines denote the result obtained by the {\it VIS\_CS} algorithm and dashed lines by {\it Clean}.}
\label{x0619}
\end{figure}
\begin{figure*}[t!]
\begin{center}
  \includegraphics*[width=0.49\textwidth]{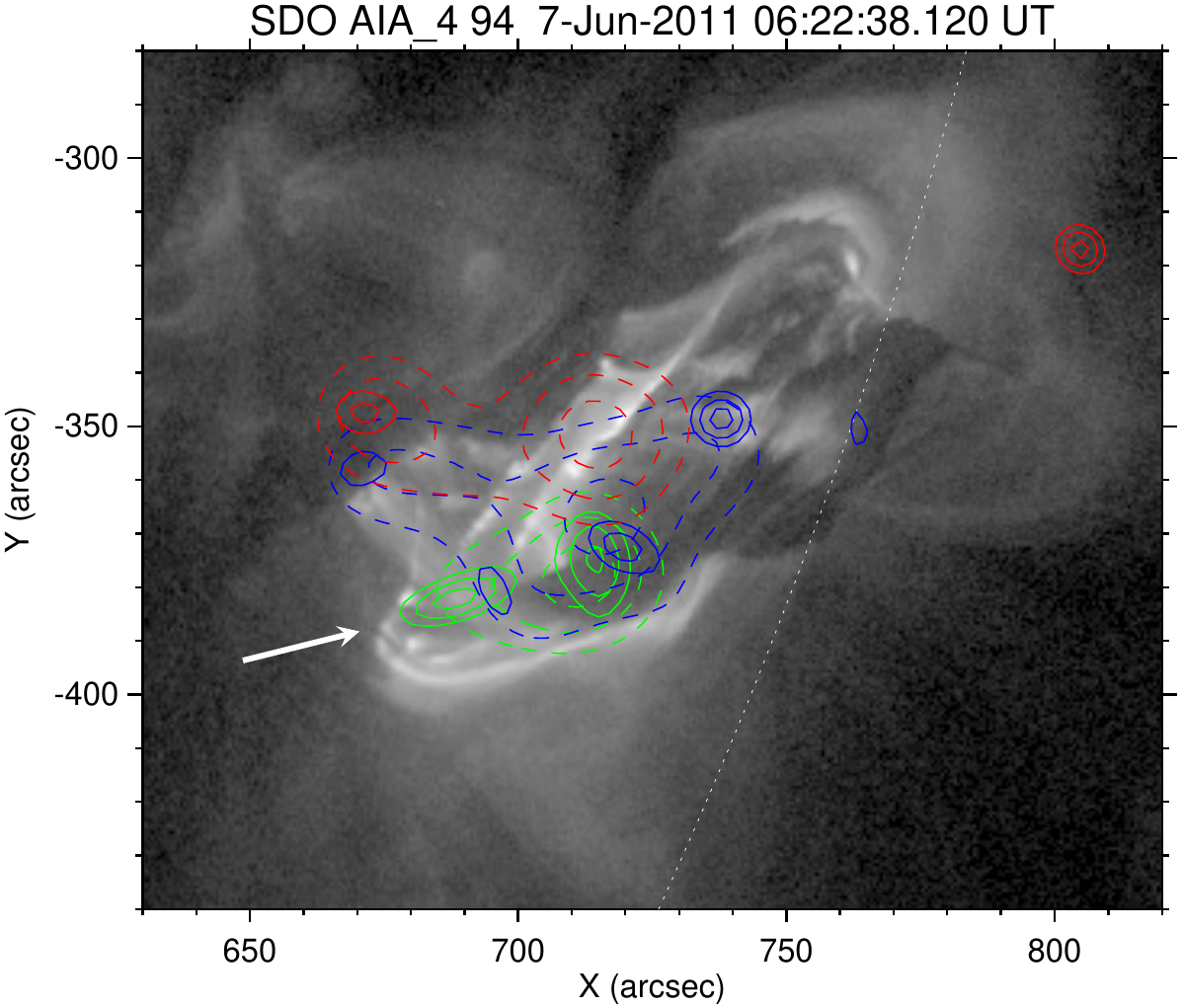}
  \includegraphics*[width=0.49\textwidth]{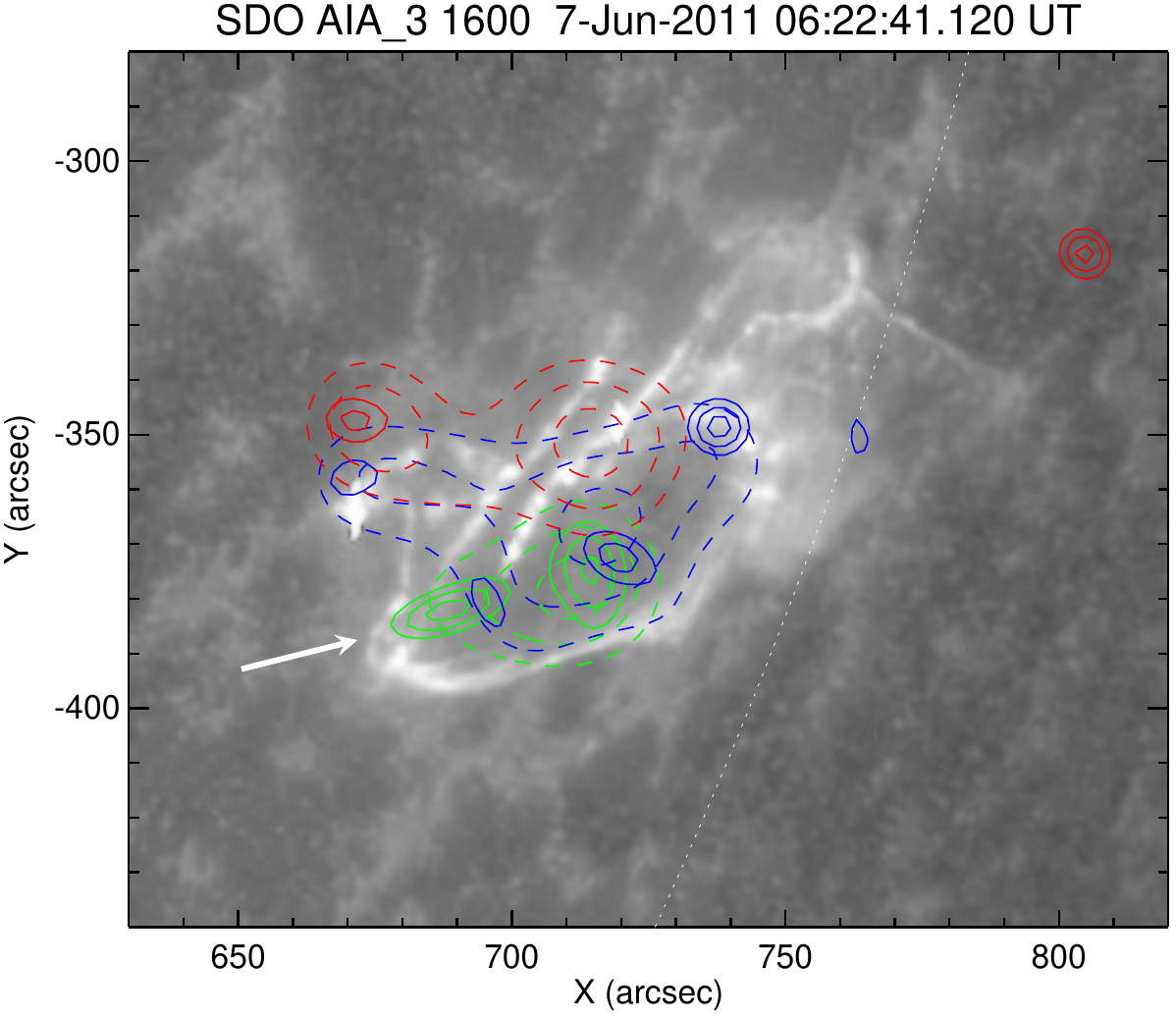}

\end{center}
    \caption{
      \textit{RHESSI} X-ray sources in the 6--12~keV (green), 12--25~keV (blue), and 25--50~keV (red) energy ranges in the
      06:22:36--06:22:48 UT time interval  superposed on the \textit{SDO}/AIA 
      94~\AA~(left) and 1600~\AA~(right)~images at 06:22:38 UT and 06:22:41 UT, respectively.
      Contours and lines styles are the same as in Fig.~\ref{x0619}.
      The white arrow points to the X-ray source and EUV brightenings at the southern footpoint of the rising filament.
      The 25--50~keV source on the right edge of the image is probably an artifact.}
\label{x0622}
\end{figure*}

\begin{figure}[h]
\begin{center}
  \includegraphics*[width=0.49\textwidth]{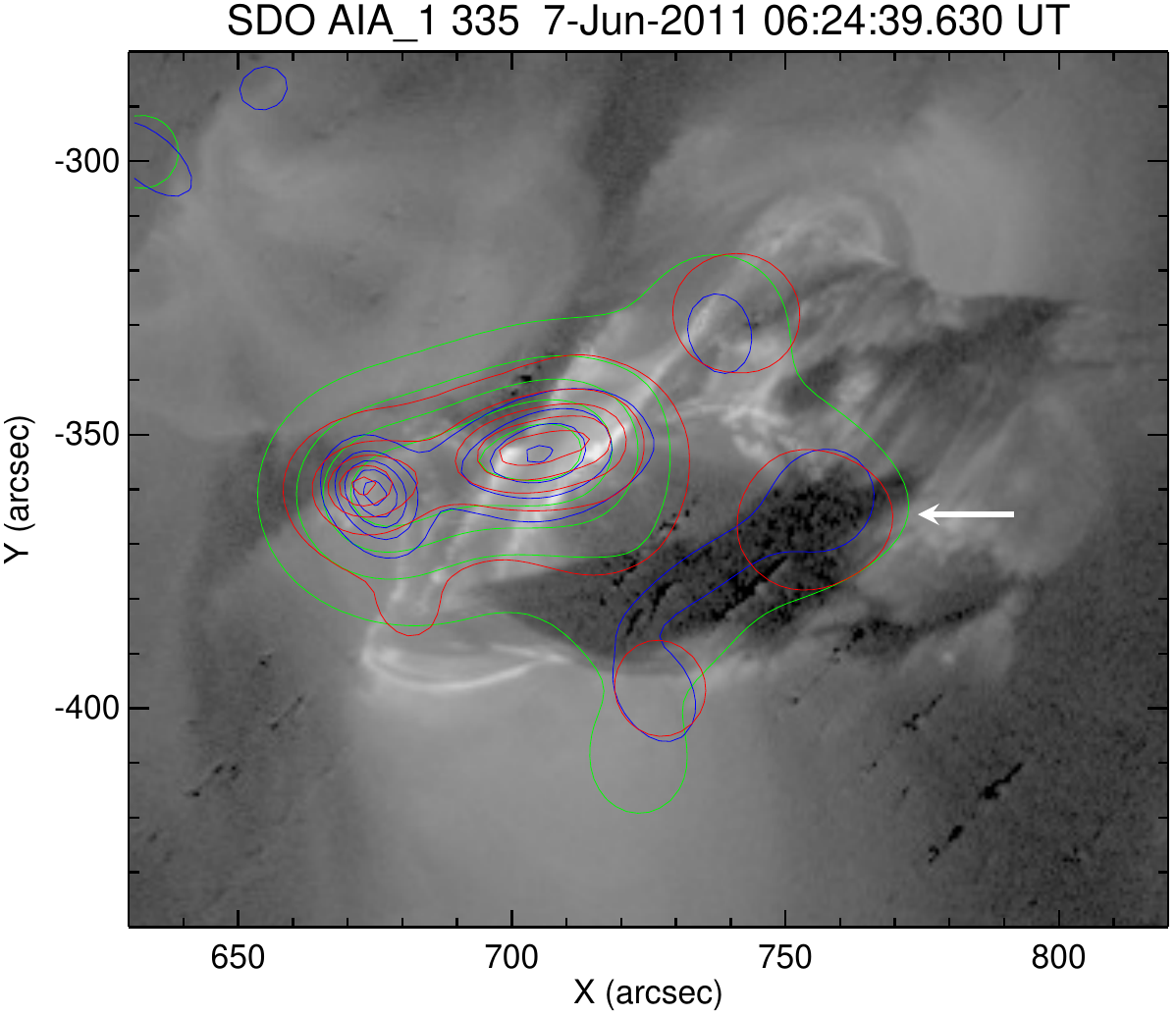}
\end{center}
    \caption{
      \textit{RHESSI} X-ray sources in the 25--50~keV range and in the 06:24:34--06:24:54 UT
      time interval
      superposed on the \textit{SDO}/AIA 335~\AA~image at 06:24:39 UT.
      Colours denote the results obtained by three algorithms: {\it Clean} (green), {\it VIS\_CS} (blue), and {\it Pixon} (red).
      Contours correspond to 10, 30, 50, 70, and 90\%  levels of the maximum in the images.
      The white arrow points to a weak source  at the upper part of the filament.
    }
\label{x0624}
\end{figure}

Figure~\ref{x0619} shows the X-ray sources (6-12 and 12-25 keV) superposed over
the \textit{SDO}/AIA 94 \AA~ image in the very early phase of the filament rise at
06:19:38 UT. At this phase the sources were located along and at the
position where at later the two parallel flare ribbons appear. A few
minutes later, the distribution of the X-ray sources changed as shown in
Figure~\ref{x0622} for 06:22:38 UT, chosen as a representative example. At this
phase the X-ray sources are transient in space and time and appear outside the
two ribbons seen in EUV emission, mainly around the southern part of the rising
filament. The arrow in this figure points to the X-ray source and EUV
brightenings at the footpoint of the rising filament (part of the rising
magnetic rope). The phase ends at the beginning of the flare impulsive phase at
06:24:15 UT as shown by the start of the DPS (Figure~\ref{blen}) and by a rapid
increase of the microwave spectrum between 06:24 and 06:25 UT
(Figure~\ref{micro}). Because the DPS is expected to be the radio emission from the
plasmoid in the magnetic rope \citep{2000A&A...360..715K}, we searched for an
X-ray source at the upper part of the rising filament. We found such a source
at energies above 12 keV and at a position of about [750,-370]~arcsec; see
the arrow in Figure~\ref{x0624}. This X-ray source is rather weak, i.e.
apparent only at the 10\% level of the maximum flux in the image, yet
was reconstructed by all three algorithms used. It occurs at
$\sim$06:24:34--06:24:54~UT, and after around $\sim$ 10~s, disappears. There
are two other sources apparent at the 10\% level, in the northward
  [720,-400] and southward [740,-330] directions, but they could be associated
  respectively with an AIA 1600~\AA\ brightening below the filament leg and an AIA brightening 
  that increases in intensity from ~ 06:24:43 UT e.g. in the 171, 193, and 211~\AA\ AIA  filters.
Because total image counts in the \textit{RHESSI} images in Figure~\ref{x0624} exceed
$10^4$, we considered all sources above 10\% and apparing in all three
algorithms used as real.

\begin{figure*}[!t]
\begin{center}
\includegraphics[width=0.98\textwidth]{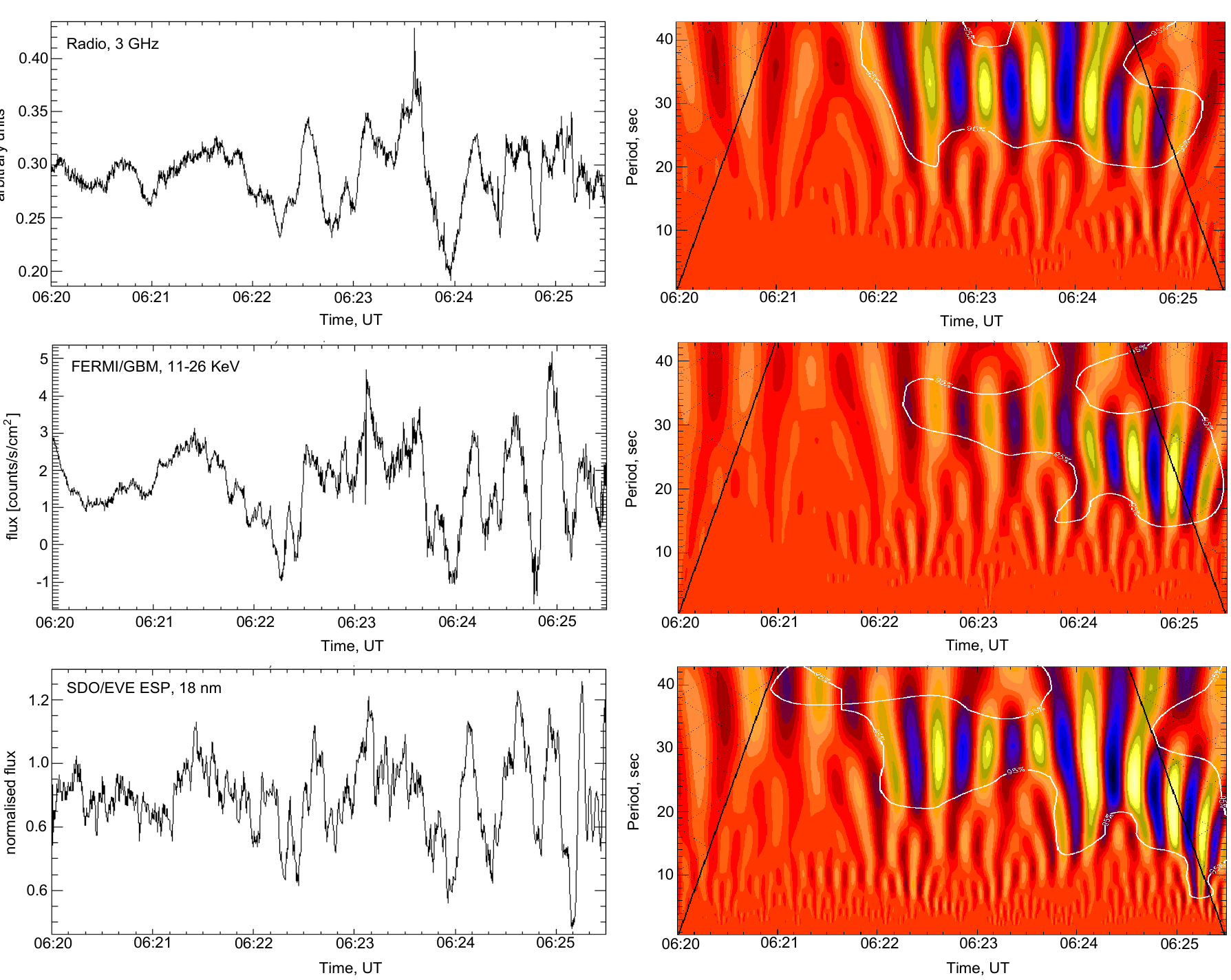}
\end{center}
\caption{Left: detrended light curves of the radio emission (3 GHz), X-ray (\textit{Fermi}/GBM, detector No. 5, 11-26 keV), and EUV (EVE/ESP, 18 nm) for
the  pre-impulsive phase of the flare,  i.e., 06:20:00--06:25:30 UT. Right: wavelet amplitude spectra corresponding to the time profiles on the left. The color background presents the signal variations in time and period. The contour lines indicate the 95\% significance level.}
\label{wav_power}
\end{figure*}

Spatially integrated X-ray spectra and their fitted parameters
  do not show any significant temporal change while
  the structures of X-ray sources and radio emission do.
  The spectra gradually become more intense and extend to higher energies.

\subsection{Spectral Processing of the Light Curves}

As already mentioned, the light curves in X-ray, EUV, and 3 GHz radio
emissions indicate quasi-periodic oscillations at the growth phase of the
flare. For spectral analysis of the observed periodicity, we used the wavelet
transform \citep{1998BAMS...79...61T}. Using this technique, we analyzed
not only the light curves presented in Figure ~\ref{light_curves}, but also the
following data: light curves observed by the two most Sun-directed \textit{Fermi}/GBM
detectors, 4 and 5, at energies above 11 keV; \textit{RHESSI} light curves in the 12--25
keV, 25--50 keV, and 50--100 keV energy ranges; curves obtained by the EVE/ESP
experiment in both the 18 nm and 26 nm bands; radio flux profiles at all
frequencies (1-35 GHz) of the Nobeyama radio polarimeters; and the correlation
curve of the Nobeyama radioheliograph. The time resolution of the 3 GHz radio
data was 0.01 s, that of the EUV EVE/ESP data was 0.25 s, and that of the X-ray \textit{Fermi} data was 0.256 s.
To compare the curves obtained with different time resolution, we reduced them
to the same resolution of 0.25 s using linear interpolation.

\begin{figure*}[!t]
\begin{center}
\includegraphics[width=0.98\textwidth]{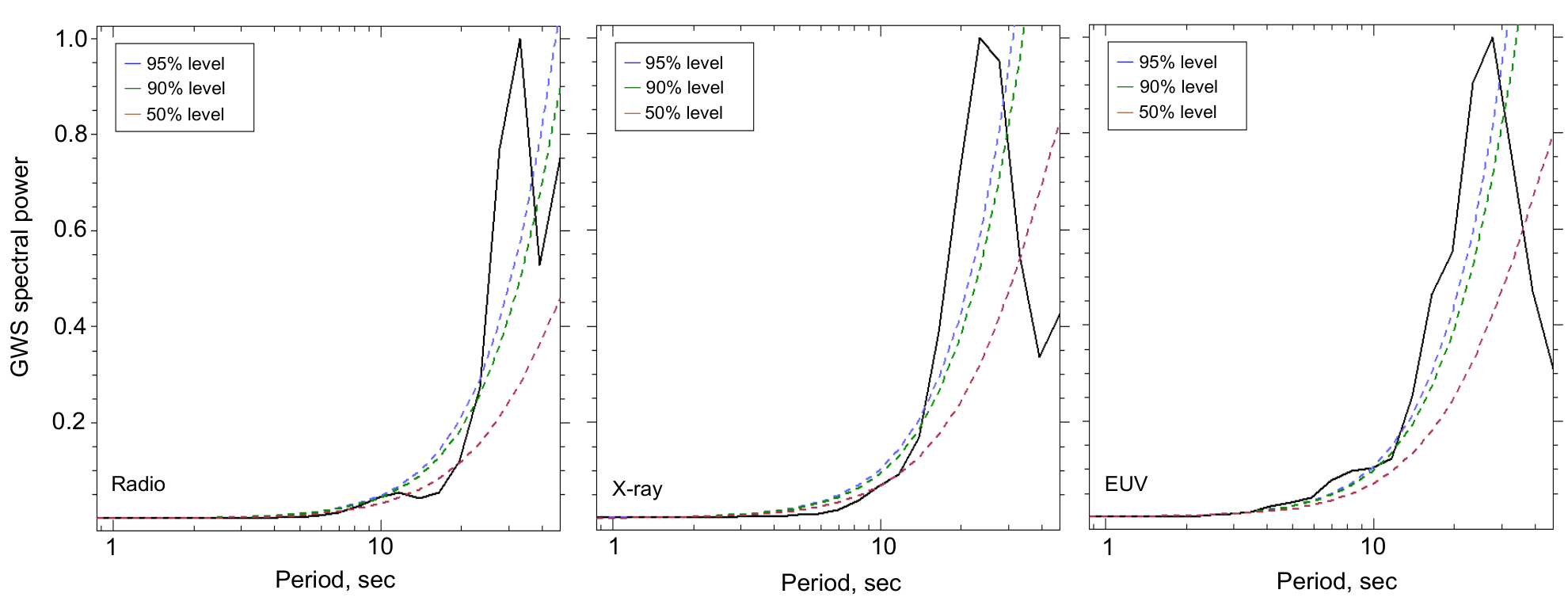}
\end{center}
\caption{Global wavelet spectra for radio, X-ray, and EUV light curves. The broken lines show the 50\% (brown),
90\% (green), and 95\% (blue) significance levels. The spectral power is normalized to the peak level of each spectrum.}
\label{wav_detail1}
\end{figure*}

In the following we present the results for light curves at radio (3
  GHz), X-ray (\textit{Fermi}/GBM, detector No. 5, 11-26 keV), and EUV (EVE/ESP, 18 nm).
In the first stage of data preparation we applied the trend removal from the light
curves using a six-degree polynomial approximation. We performed a
least-squares polynomial fit of the curves with matrix inversion and optional
weighting to determine the fitting coefficients. This allowed us to obtain the
detrended curves in the 06:20:00--06:25:30 UT time interval
(Fig.~\ref{wav_power}, ~left panels). We can see that the signal variance and
the corresponding signal-to-noise ratio increase and reach a maximum
before the flux peak around 06:25:30 UT. The noise level is insignificant
compared to the oscillations, and we can ignore its influence on the signal
level. When comparing the profiles, it can be seen that their extremes are
almost in phase.

To study the spectral structure of the observed oscillations, we
constructed the amplitude wavelet spectra (Fig.~\ref{wav_power}, ~right panels)
and superimposed the contour with the 95\% significance level. To determine the
reliability of the found periodicity, we compared them with the background
spectra adopting a power-law noise assumption. Then, using a chi-squared
distribution we found the 95\% significance interval. In our work we applied
the software previously used for the pixelised wavelet filtering spectral
analysis \citep{2002ChJAA...2..183S, 2008SoPh..248..395S,
2010SoPh..261..281K}.

We can see that for the 95\% significance level there is a periodicity with
an average value near 30~s. There is a drift toward shorter
periods with increasing oscillation power. The half-periods of the
oscillations, displayed as bright and dark patches, almost coincide in time for
the different energy ranges (radio, X-ray, EUV). To obtain a more accurate value of
the oscillation periods, we calculated the global wavelet spectra and
estimated the 50\%, 90\%, and 95\% significance levels
(Fig.~\ref{wav_detail1}). The obtained peaks are above the 95\% significance
level and correspond to periods of $\sim$ 33~s (radio), $\sim$ 25~s
(X-ray), and $\sim$ 28~s (EUV).

To understand how the flare light curves are interconnected at
different emissions, we used cross-wavelet analysis and wavelet coherence
\citep{1998BAMS...79...61T, 1999JCli...12.2679T}. These spectral techniques are
used to analyze the coherence, cross-correlation, and phase between signals as
functions of time and frequency, obtained from the wavelet spectra of two
signals. The cross-spectrum is determined by multiplying the wavelet power spectrum
of the one signal with the complex conjugate wavelet spectrum of the other. The
maximum values of cross-correlation are represented in period-time
coordinates and depend on the correlation of oscillation power at certain
frequencies. Another method of spectral analysis of two profiles is to
construct a wavelet coherence spectrum. This spectrum represents localized
amplitudes and phases of correlation wavelet coefficients in the time--frequency
domain. The value of the spectra averaged over all times gives us the global
coherence and global phase depending on the period. Coherence ranges from 0 to
1, where 0 is the absence of coherence, and 1 is excellent coherence. The
global phase range varies from $180^\circ$ to $-180^\circ$.

\begin{figure*}[!t]
\begin{center}
\includegraphics[width=0.98\textwidth]{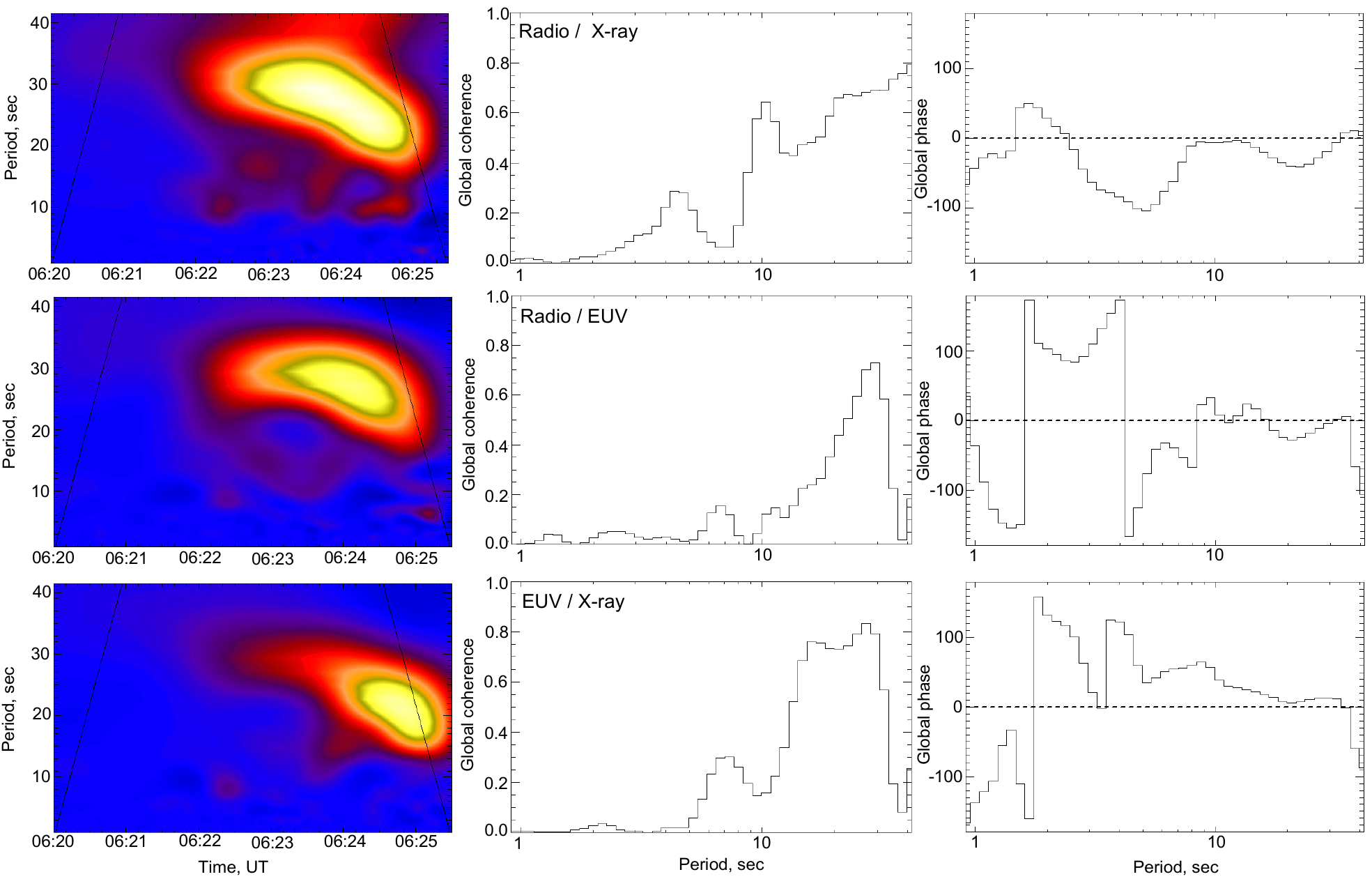}
\end{center}
\caption{Wavelet cross-correlation spectra (left), global coherence profiles (center),
and global  phase profiles (right) for each pair of the light curves: radio--X-ray, radio--EUV, and EUV--X-ray.
The horizontal broken lines show the zero level of the global phase. The coherence is from 0 to 1 and phase  from $-180^\circ$ to $180^\circ$.}
\label{wavelets}
\end{figure*}
  
We prepared the wavelet cross spectra for each pair of the signals
(radio--X-ray, radio--EUV, EUV--X-ray) and calculated their global coherences and
global phases.  It can be seen that (Fig.~\ref{wavelets}, ~left panels),  for
all pairs, the power of cross-oscillations is maximal (cross-correlated) near
the $\sim$30~s period. Also, simultaneous period drifts toward shorter
periods can be seen. This pattern we noted earlier in the analysis of the
individual light curves (Fig.~\ref{wav_power}). Global coherence profiles show
a significant level of $\sim$ 0.7 - 0.8 for the 30~s periodicity
(Fig.~\ref{wavelets}, ~central panels). The averaged phase over time of these
oscillations is near zero, shown by the broken line, indicating that the signals
are in phase (Fig.~\ref{wavelets}, ~right panels).

The $\sim$ 30 s period was also recognized in the \textit{RHESSI} channels in the 12--25 and 25--50 keV, \textit{Fermi}/GBM 26--50 keV channels from detectors 4 and 5, in the 26 nm EVE/ESP channel, and in the radio frequencies 2 and 3.75 GHz of the Nobeyama radio polarimeters. Some indication of this period was also found in the correlation curve of the Nobeyama radioheliograph. The found $\sim$ 30~s periodicity with probability 95\% for independent instruments indicates the reality of these oscillations. We also searched for the $\sim$ 30 s period at lower radio frequencies, i. e. below 2 GHz, but without result.

In the wavelet amplitude spectra in Figures~\ref{wav_power}~(right)
some vertically long and short striations appear. While the vertically
long striations correspond to a very broad interval of periods (therefore no
oscillation), the vertically short striations could indicate some further
oscillations. But because these further possible oscillations were recognized
only on a few wavelet amplitude spectra, we did not study them in detail. Only
the $\sim$ 30 s period was found in nearly all analyzed wavelet amplitude
spectra.

In Figure~\ref{blen} we show also the DPS. Although this DPS could be analyzed as
that in \cite{2018SoPh..293...62K}, due to the relatively low temporal resolution
(0.25 s) we did not analyze periods in the DPS in detail. As shown in
\cite{2018SoPh..293...62K}, the typical periods in DPSs are about 1~s or even
fractions of a second. Nevertheless, in the present DPS, the periods in the
4--8 s range were found.

\section{Discussion and Conclusions}

\begin{figure}[t]
\begin{center}
\includegraphics*[width=0.48\textwidth]{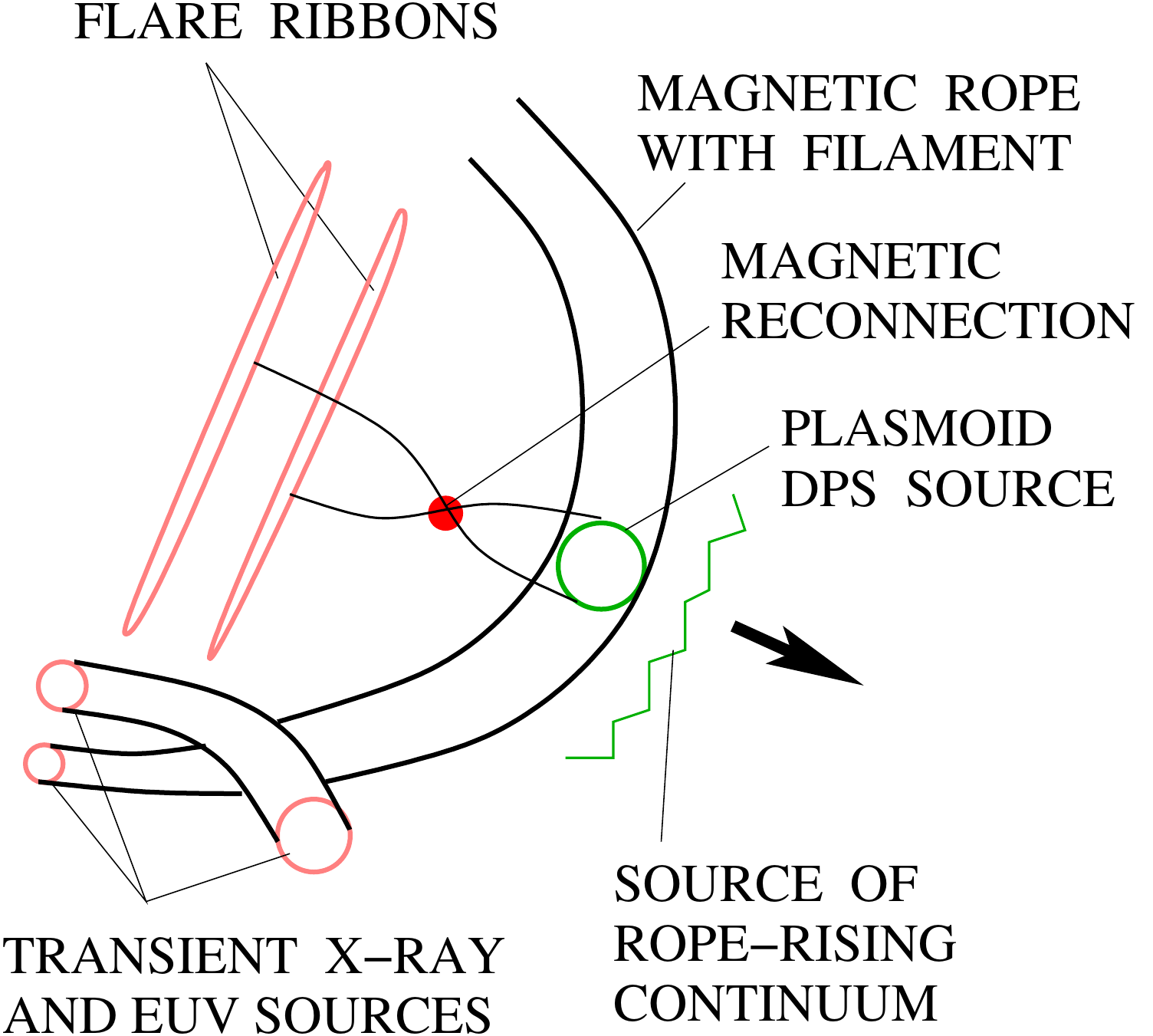}
\end{center}
    \caption{Scenario of processes during the filament rise in the 2011 June 7 flare.
    In the pre-impulsive phase the rope-rising continuum and
transient X-ray and {\bf E}UV sources were generated by the magnetic reconnection
between the rising magnetic rope and above-lying magnetic fields as well as
by the magnetic reconnection inside the magnetic rope. The impulsive phase
started with magnetic reconnection below the rising filament and by the
appearance of the DPS and the X-source (plasmoid) at the upper part of the rising
filament.}
\label{scenario}
\end{figure}

We present an unusual radio continuum drifting from higher to lower
frequencies, observed during a filament rise. As we already mentioned, to our
knowledge only one similar emission has been observed previously, i.e. at the
beginning of the 2001 September 24 flare. We call this continuum the
rope-rising continuum. Assuming that this continuum is generated by a plasma
emission mechanism on the fundamental frequency, and using the Aschwanden's
model of the solar atmosphere~\citep{2002SSRv..101....1A}, the velocity of the
agent generating the starting boundary of this continuum is about
400 km s$^{-1}$. This velocity is comparable to that of the rising
filament (200 - 400 km s$^{-1}$ ~\citep{2012ApJ...745L...5C}). Therefore, we
propose that this continuum is generated by the rising magnetic rope in which
the filament is embedded. The magnetic rope interacts with the magnetic field
above it, and during this interaction (localized reconnections), electrons
are accelerated and produce this unusual continuum; see the scenario of flare
processes in the pre-impulsive phase in Figure~\ref{scenario}. We also
note that the upper boundary of the magnetic rope, where the interaction with
the above-lying loops takes place, can move upward faster than the filament,
owing to the magnetic rope expansion.

We also present a DPS observed on unusually
low frequencies (220--450 MHz). Usually they are observed on much higher
frequencies in the GHz frequency range~\citep{2015ApJ...799..126N}. It seems
that this is connected with the eruption of an unusually huge filament. In this
case the magnetic reconnection generating the plasmoid and DPS started when
the upper part of the magnetic rope, carrying the filament, reached a very high
coronal altitude. Because the plasmoid density depends on that of the
surrounding plasma, it was lower than usual  and
thus the plasma emission frequency from the plasmoid (DPS frequency) was
unusually low.

As concerns the X-ray sources, we found that at the very beginning of the
filament rise the X-ray sources were at a position where later two parallel ribbons
appeared in the EUV emission. Then the X-ray
sources were transient in space and time and also extended out of the position
of these parallel ribbons. We think that the appearance of transient X-ray
source located out of the parallel ribbons is also connected with the interaction of
the rising magnetic rope and the above-lying magnetic field. The accelerated
electrons propagate from the region of interaction along different magnetic
field lines downward to different locations, where they heat the plasma
(emitting in UV and EUV) and produce transient X-ray emission by
thermal and non-thermal bremsstrahlung. Signatures of the complex structure of
the above-lying magnetic field can be seen, e.g., in Figure~\ref{x0619} as 
EUV emission loops perpendicular to the filament with their footpoints at
$\sim$ [680, -360] arcsec.

The phase related to these transient X-ray sources ended during the flare
impulsive phase shown by the DPS starting at 06:24:15 UT. The DPS was associated
with a sudden increase in the microwave emission and observation of the 25-50
keV X-ray source in the upper part of the rising filament (see the arrow in
Figure~\ref{x0624}). We interpret this X-ray source as emission from the
plasmoid (Figure~\ref{scenario}), which appears as a result of the magnetic
reconnection located below the rising filament. But this X-ray source
disappeared after a few seconds. We think that this disappearance is
by the intensity decrease of this source due to the plasma density
decrease in an expanding filament and/or by the intensity increase of other
X-ray sources and the relatively low dynamic range of \textit{RHESSI}. 
After the beginning of the flare impulsive phase, i.e. after 06:26 UT, the
X-ray sources appeared at the parallel flare ribbons \citep{2013ApJ...777...30I},
as usually observed.

In this pre-impulsive flare phase we also found EUV brightenings and the X-ray
6--12~keV source close to the filament footpoint. This agrees with the
observation of the hot magnetic rope during the pre-impulsive stage of an
eruptive 2012 July 19 flare \citep{2016ApJ...820L..29W}. These observations
indicate that the plasma is heated and probably also superthermal particles are
accelerated in the magnetic rope where the filament is embedded. Such 
heating and acceleration can be due to the magnetic reconnection inside the
magnetic rope where the helical magnetic field is expected. During the rising
of this magnetic rope, owing to its kinking, internal reconnection can be
triggered. This type of the reconnection can be added to the list of possible
types of magnetic reconnection in eruptive flares presented by
\cite{2019A&A...621A..72A} as the rr--rr type.

During the filament rise at about 06:22--06:25 UT we detected
$\sim$30~s quasi-periodic oscillations simultaneously in hard X-ray, EUV, and
2, 3, 3.75 GHz radio emissions. In global wavelet spectra, peaks in
periods above the 95\% significance level were found at $\sim$33~s (radio),
$\sim$25~s (X-ray), and $\sim$28~s (EUV). There was also a periodicity drift
toward shorter periods with increasing oscillation power. The
half-periods of oscillations displayed as bright and dark patches almost
coincided in time for different energy ranges (radio, X-ray, EUV). Furthermore,
the global coherence profiles showed a significant level, $\sim$~0.7--0.8, for
the $\sim$30~s periodicity and the global phase over time of
these oscillations was near zero, indicating that the signals with the $\sim$30
s period in radio, X-ray, and EUV are in phase.

Oscillations in this flare phase with similar periods are known; see e.g.,
\citet{2003SoPh..218..183F} and \citet{2014ApJ...791...44H}. These oscillations can
be caused by fast kink mode oscillations as proposed by
\cite{2014ApJ...791...44H} or by an oscillatory acceleration/injection process
caused by oscillation of the magnetic loop interacting with the rising magnetic
rope where the filament is embedded ~\citep{2009SSRv..149..119N}. Through
these interactions (magnetic reconnections) electrons are accelerated and
cause the quasi-periodic variations in the observed hard X-ray, EUV, and
radio gyro-synchrotron emission. On lower radio frequencies (below 2 GHz),
where bursts are preferentially generated by the plasma emission mechanism, no
such period was found. It appears that oscillations end due to a much stronger
process (magnetic reconnection below the rising filament) in the flare
impulsive phase.

Besides the above-mentioned explanations for the oscillation, there is
a further possibility: after the interaction of the rising magnetic
rope with the above-lying loops, some of the latter are reformed into
loops below the rising magnetic rope. During this process a loop starts to
oscillate, and moreover, this loop is shrinking. Its period of oscillation can
be expressed in the simplest way as
\begin{equation}
P \sim \frac{L}{v},
\end{equation}
where $L$ is the loop length and $v$ is the characteristic perturbation speed.
In our case we assume that $v = v_A$, where $v_A = B/\sqrt{4\pi \rho}$ is the
Alfv\'en velocity, $B$ is the magnetic field, and $\rho$ is the plasma density.
The plasma density and magnetic field change during the loop shrinking;
therefore, for the oscillation period we can write
\begin{equation}
P \sim \frac{\sqrt{4 \pi M V}}{F},
\end{equation}
where $M = \rho V$ is the total plasma mass in the shrinking loop, $V = L S$ is
the loop volume, $S$ is the loop cross-section area, and $F = B S$ is the magnetic flux.
When we neglect the plasma evaporation  in this loop, then $M$ and $F$
are conserved during the loop shrinking. Thus, the period of oscillation
decreases with the loop volume as $P \sim \sqrt{V}$. This
interpretation of the observed oscillation is supported by the detected
shortening of the oscillation period from about 30~s to 20~s; see
Figure~\ref{wav_power}. In this interpretation it means that the starting loop volume
$V_{\text{start}}$ shrinks to the 0.44 $\times V_{\text{start}}$ at the end of
the oscillation. Furthermore, during the loop shrinking the plasma density
and magnetic field increase, and so the radio gyro-synchrotron, X-ray,
and EUV emissions also increase, as observed. However, verification of this
explanation needs further observations and analysis.

Figure~\ref{scenario} summarizes our interpretation of observations of the 2011
June 7 flare. In the pre-impulsive flare phase the rope-rising radio
continuum and transient X-ray and {\bf E}UV sources were generated by  magnetic
reconnection between the rising magnetic rope and above-lying magnetic fields
as well as by magnetic reconnection inside the magnetic rope. The impulsive
phase started with magnetic reconnection below the rising filament and by the
appearance of a DPS and the X-source (plasmoid) at the upper part of the
rising filament.

\acknowledgements The authors thank the referee for comments that improved the
paper.  M.K. and J.K. acknowledge support from the project RVO:67985815 and GA
\v{C}R grants 17-16447S, 18-09072S, and 19-09489S. The study was
performed within the basic funding from FR program II.16, RAS program KP19-270,
and partially supported by the Russian Foundation for Basic Research (RFBR)
under Grant 17-52-80064 BRICS-a. We thank Dr. C. Monstein for the
Callisto spectrum, J. Dud\'{\i}k for his help and advice with AIA data, J.
Ryb\'ak for the help with the radio spectrum, \textit{RHESSI} software and Nobeyama
teams. We also acknowledge the use of the \textit{Fermi} Solar Flare Observations
facility funded by the \textit{Fermi} GI program
(\url{http://hesperia.gsfc.nasa.gov/fermi_solar/}). AIA data are courtesy of
NASA/\textit{SDO} and the AIA science team.

\end{document}